\documentclass[epj,final]{svjour}

\usepackage{graphics}

\begin{document}
\title{Reliable teleportation in trapped ions}
\author{E. Solano\inst{1}\inst{2}, C. L. Cesar\inst{1}, R. L. de Matos Filho\inst{1},
and N. Zagury\inst{1}}

\institute{Instituto de F\'{\i}sica, Universidade Federal do Rio
de Janeiro, Caixa Postal 68528, 21945-970 Rio de Janeiro, RJ,
Brazil \and Secci\'{o}n F\'{\i}sica, Departamento de Ciencias,
Pontificia Universidad Cat\'{o}lica del Per\'{u}, Apartado 1761,
Lima, Peru}

\date{\today}

\abstract{
We study a method for the implementation of a reliable
teleportation  protocol (theoretically, $100\%$ of success) of
internal states in trapped ions. The generation of the quantum
channel (any of four Bell states) may be done respecting technical
limitations on individual addressing and without claiming the
Lamb-Dicke regime. An adequate Bell analyzer, that transforms
unitarily the Bell basis into a completely disentangled one, is
considered. Probable sources of error and fidelity estimations of
the teleportation process are studied. Finally, we discuss
experimental issues, proposing a scenario in which the present
scheme could be implemented.
}

\maketitle

\section{Introduction}

Entanglement, or non separability, is a natural consequence of linearity in
quantum mechanics and its careful study is one of the relevant tasks in
fundamentals of physics. Since Einstein-Podolski-Rosen seminal work
\cite{EPR}, non local effects in two, or more, entangled particles have been
the object of intense research. Local hidden variables theories, partially
designed to avoid non intuitive explanations, led to results in contradiction
with several experiments \cite{bell,aspect}. As a consequence, the
deterministic generation of entanglement in a pair of two-level particles,
being this entangled bipartite system the simplest one, has become an
important theoretical and experimental demand. In particular, there exists a
fundamental interest in the generation of Bell states, a complete basis in a
four dimensional Hilbert space made out of maximally entangled orthogonal
states. Additionally, it has been shown that producing Bell states in two
macroscopically distant particles is of extreme importance to experimentalists
for the implementation of quantum teleportation
~\cite{bennett,exptelep1,exptelep2}, quantum cryptography~\cite{crypto} and
quantum computation~\cite{QC,cirac1}.

Massive particles, like atoms or ions, are good for storing
quantum information and photons are the natural messengers for
communicating them. Controlled entanglement between massive
particles has been achieved in the case of atoms crossing a
high~$Q$ cavity\cite{haroche,englert} and in the case of ions in a
trap \cite{turchette,sackett}.

Bell states provides the essential non local ingredient for performing quantum
teleportation, where a disembodied transmission of an arbitrary state, from
one system to another located at a remote place, is made through a classical
and a quantum channel.  After the original proposal of quantum
teleportation~\cite{bennett} (hereafter called BBCJPW), many experimental
efforts have been made in order to test teleportation of an arbitrary state of
discrete~\cite{exptelep1} or continuous variables~\cite{exptelep2}. In all
these experiments the particles involved were photons. The main difficulty for
the implementation of the BBCJPW protocol for teleportation of discrete states
is the complete measurement of a non-degenerate Bell
operator~\cite{braunstein}. This would require the projection of an input
two-particle state on one of four maximally entangled orthogonal states and
the production of a unique signal for each one. In fact, it has recently been
shown that, in the case of teleportation of discrete variables, it is not
possible to perform a complete Bell-operator measurement without an effective
quantum interaction between the involved particles~\cite{vaidman,lutkenhaus}.
Due to the lack of a photon-photon interaction, it seems impossible to obtain
a reliable teleportation of photon polarization states that follows closely
the original BBCJPW protocol. For this reason, it seems more promising to look
for systems where effective quantum interactions between their components may
be implemented more easily, like trapped ions or systems composed of atoms
interacting with electromagnetic cavities~\cite{davidovich,cirac2}.
Teleportation with trapped ions has an additional important advantage: we are
dealing with a single long living quantum system, confined into a very small
region of space and that remains at our disposal for further manipulation. In
this paper we propose a method for implementing a reliable teleportation
protocol (theoretically, $100 \%$ of success) of an arbitrary internal state
in trapped ions.

In Section II, we discuss a proposal for reliable teleportation of
arbitrary two-level electronic states between ions kept in two
well separated traps. This is accomplished by means of a Bell
state as the quantum channel and the implementation of an adequate
Bell analyzer using well determined unitary operations and
projections, avoiding undesired entanglement of the internal
states with the motional degrees of freedom.  Due to the use of
ion-ion quantum interaction, all the measurements needed in this
scheme are performed exclusively on single ions. In this way, our
Bell analyzer distinguishes efficiently the four Bell states.
Numerical estimations of the fidelity of the teleportation process
are also presented. In Section III, we consider some
generalization of our scheme for implementing entanglement
teleportation and entanglement swapping. In Section IV, we discuss
relevant experimental issues for the practical realization of
 our teleportation scheme. Finally, we summarize our main
results in Section V.

\section{Teleportation}
\subsection{Scheme}

Our procedure, along the lines of the BBCJPW protocol
\cite{bennett}, can be sketched as follows (see Fig.~1): Trap A
(Alice station) contains ion 1 with an arbitrary electronic state,
$\alpha \left| \downarrow_1 \right\rangle +\beta \left| \uparrow_1
\right\rangle$ and trap $B$ (Bob station) contains ions 2 and 3 in
a maximally entangled state, e.g. any of the four Bell states. Ion
2 is then transferred adiabatically to Alice station, who performs
a suitable operation on the joint Hilbert space of ions 1 and 2
transforming unitarily the Bell base into a disentangled base.
Subsequently, Alice measures their individual electronic states,
completing the so called Bell analyzer. Then, Alice informs to Bob
the result of her measurements, consisting in two bits of
classical information. Bob uses them to perform one, out of four,
specific unitary transformation on ion 3, whose electronic state
is left at the original state of ion 1.

\begin{figure}[htb]
\centerline{\resizebox{0.35\textwidth}{!}{
\includegraphics{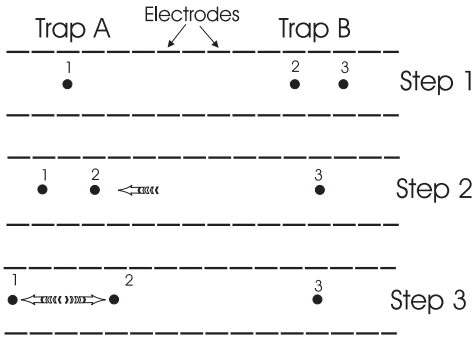}} }
\vspace{1.5cm} \caption{Sketch of the trap configuration for
teleportation of the electronic state of ion 1 to ion 3. In the
first step, ion 1 has an arbitrary electronic state, while ions 2
and 3 are in an Bell state. In step 2, the second ion is
transported to trap $A$, where it undergoes entanglement with ion
1. As a last step, ions 1 and 2 are separated so that their
individual states can be measured via fluorescence.} \label{fig1}
\end{figure}

\subsection{Quantum channel}

The first task in the implementation of our teleportation scheme
is the preparation of the quantum channel, i.e., the deterministic
generation of an electronic Bell state in trap B. In a real
situation we will deal with initial motional thermal states, where
contributions of different vibrational Fock states, associated
with the center of mass mode of frequency $\nu$ and with the
stretch mode of frequency $\nu_r$, are relevant. Therefore, we
will consider the general proposal of Ref. \cite{SMZ} for the
generation of the four Bell states. According to it, one of the
Bell states may be generated, with present techniques and not
claiming the Lamb-Dicke regime, through the use of two pairs of
Raman lasers acting on {\it both} ions. The two Raman laser pairs
induce dispersive interactions between two long-living hyperfine
levels, $|\downarrow\rangle$ and $|\uparrow\rangle$ (energy
difference equal to $\hbar\omega_0$). One pair produces a
dispersive transition to the $k^{th}$ blue sideband
($\omega_{I}=\omega _{o}+k\nu-\delta$) while the other causes a
dispersive transition to the $k^{th}$ red sideband
($\omega_{II}=\omega_{o}-k\nu+\delta$). See Fig. 2 for a sketch of
the energy level diagram.

\vspace*{1cm}

\begin{figure}[htb]
\centerline{\resizebox{0.25\textwidth}{!}{
\includegraphics{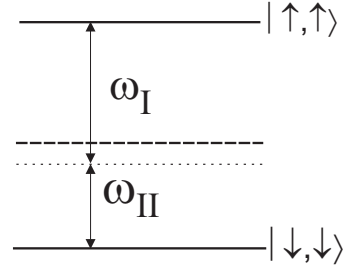}} }
\vspace*{0.5cm} \caption{Energy level diagram. A stimulated
electronic transition is induced by means of two dispersive laser
excitations with frequencies $\omega_{I}=\omega _{o}+k\nu-\delta$
and $\omega_{II}=\omega_{o}-k\nu+\delta$ in such a way that
$\omega_{I}+\omega_{II}=2\omega_{o}$. } \label{fig2}
\end{figure}

This process is described by the general effective Hamiltonian~\cite{SMZ}
\begin{eqnarray}
\widehat{H}_{\rm eff}&&=\hbar\Omega _{k}\lbrack
\widehat{S}_{+j}\widehat{S}_{+m}e^{2i\phi }+ (-1)^k\left(
\widehat{S}_{+j}\widehat{S}_{-m}e^{i\phi_{o}}+\frac{1}{2}\right)
\rbrack   \nonumber \\ &&\times \lbrack
\widehat{F}_{k}^{2}(\hat{n}-k,\hat{n}_{r}) \frac{\hat{n}!} {\left(
\hat{n}-k\right) !}-
\widehat{F}_{k}^{2}(\hat{n},\hat{n}_{r})\frac{\left( \hat{n}
+k\right) !}{\hat{n}!}\rbrack \nonumber \\ &&+{\rm H}.c.
\label{Heff}
\end{eqnarray}
where $\Omega _{k}=2 |\Omega |^{2}\left( i\eta \right)
^{2k}/\delta$. Here, $\Omega$ is the Rabi frequency associated
with each Raman pair and $\eta$ is the CM Lamb-Dicke parameter.
$\widehat{S}_{+m}=|\!\!\uparrow_m\rangle\langle\downarrow_m\!\!|$
and $\widehat{S}_{-m}=S_{+m}^{\dagger}$ are the electronic raising
and lowering operators acting on ion $m$, $\phi$ is the effective
phase, taking as equal, of each Raman laser pair and $\phi_{0}$ is
the known phase difference associated with the equilibrium
separation of the two ions. $\hat{n}$ and $\hat{n_r}$ are the
harmonic oscillator number operators associated with the center of
mass (CM) and the relative modes, respectively. Also,
\begin{equation}
\widehat{F}_{k}\left( \hat{n},\hat{n}_{r}\right) =\sum f_{k}(n,
n_{r})|n,n_{r}\rangle \langle n,n_{r}| \label{efe1}
\end{equation}
with
\begin{equation}
f_{k}(n,n_r)=e^{-(\eta ^{2}+\eta_r^{2})/2}\frac{n!}{(n+k)!}
L_{n}^{k}(\eta ^{2})L_{n_r}^{0}(\eta _{r}^{2}),  \label{efe2}
\end{equation}
where
\begin{eqnarray}
L_{m}^{k}(x)=\sum^{m}_{\ell=0}(-1)^{\ell} \pmatrix{m+k \cr
m-\ell\cr} \frac{x^{\ell}}{{\ell}!}
\end{eqnarray}
are associated Laguerre polynomials. The first term of Eq.~(\ref{Heff}) and its Hermitian
conjugate, acting only in the subspace determined by the states $\lbrace |\!\!
\downarrow \downarrow\rangle,|\!\! \uparrow \uparrow \rangle \rbrace$,
describe two-photon processes leading to the simultaneous excitation or
deexcitation of the electronic states of the two ions.  The second term and
its Hermitian conjugate, acting exclusively in the subspace spanned by the
states $\lbrace |\!\!\downarrow \uparrow \rangle, |\!\! \uparrow \downarrow
\rangle \rbrace$, describe processes where one ion undergoes a transition from
the ground to the excited electronic state and the other ion makes a
transition in the inverse direction, both processes taking place
simultaneously.  The third term  is a self energy term where the
energy shifts in the two effective ionic levels are equal and in the same
direction, just producing a time-dependent overall phase in the states
evolution, as we will see below. The vibrational part of this Hamiltonian,
completely factorized on the right, contains only number operators that will
not change the population of a two mode Fock state basis $|n,n_r\rangle$. The
first (second) term of the vibrational part has its origin in a virtual two
step process: in the first one, $k$ vibrational quanta are annihilated
(created) and in the second one, the same number of vibrational quanta are
created (annihilated), preserving the number of motional quanta. It is
important to note that, when the vibrational part of the initial state is the
ground state $(n=0)$, the first term of the vibrational part vanishes due to
the fact that, in this case, is impossible to annihilate a motional quantum.
We may also notice that there is no effective coupling when $k=0$, that is,
when no collective mode is excited together with the intermediate virtual
electronic levels.

In particular, we are interested in the temporal evolution of
initial states of the form $\left| \downarrow ,\downarrow
\right\rangle _{n,n_{r}}\equiv \left| \downarrow,\downarrow
\right\rangle \otimes \left| n,n_{r}\right\rangle$,$\left|
\uparrow, \uparrow \right\rangle_{n,n_r}$, $\left| \downarrow,
\uparrow \right\rangle_{n,n_r}$ and $\left| \uparrow, \downarrow
\right\rangle_{n,n_r}$. So, for an interaction time $t$ we have
\begin{eqnarray}
e^{-\frac{i}{\hbar}\widehat{H}_{\rm eff}t}\left| \downarrow,
\downarrow \right\rangle_{n,n_r} =&& e^{-i(-1)^{k}\Omega
_{nn_{r}}^{k}t} \lbrack \cos ( |\Omega _{nn_{r}}^{k}|\,t) \left|
\downarrow, \downarrow \right\rangle \nonumber \\ &&+
i(-1)^{k}e^{2i\phi }\sin ( |\Omega _{nn_{r}}^{k}|\,t) \left|
\uparrow, \uparrow \right\rangle \rbrack_{n,n_r} \nonumber \\
e^{-\frac{i}{\hbar}\widehat{H}_{\rm eff}t}\left| \uparrow,
\uparrow \right\rangle_{n,n_r} =&& e^{-i(-1)^{k}\Omega
_{nn_{r}}^{k}t} \lbrack \cos ( |\Omega _{nn_{r}}^{k}|\,t) \left|
\uparrow, \uparrow \right\rangle \nonumber \\ &&+
i(-1)^{k}e^{-2i\phi}\sin ( |\Omega _{nn_{r}}^{k}|\,t) \left|
\downarrow, \downarrow \right\rangle \rbrack_{n,n_r} \nonumber \\
e^{-\frac{i}{\hbar}\widehat{H}_{\rm eff}t}\left| \downarrow,
\uparrow \right\rangle_{n,n_r}=&& e^{-i(-1)^{k}\Omega
_{nn_{r}}^{k}t} \lbrack \cos ( |\Omega _{nn_{r}}^{k}|\,t) \left|
\downarrow, \uparrow \right\rangle \nonumber \\ &&+
ie^{i\phi_o}\sin ( |\Omega _{nn_{r}}^{k}|\,t) \left| \uparrow,
\downarrow \right\rangle \rbrack_{n,n_r} \nonumber \\
e^{-\frac{i}{\hbar}\widehat{H}_{\rm eff}t}\left| \uparrow,
\downarrow \right\rangle_{n,n_r}=&& e^{-i(-1)^{k}\Omega
_{nn_{r}}^{k}t} \lbrack \cos ( |\Omega _{nn_{r}}^{k}|\,t) \left|
\uparrow, \downarrow \right\rangle \nonumber \\&& +
ie^{-i\phi_o}\sin ( |\Omega _{nn_{r}}^{k}|\,t) \left| \downarrow,
\uparrow \right\rangle \rbrack_{n,n_r} , \label{Evols}
\end{eqnarray}
where
\begin{eqnarray}
\Omega  _{n n_{r}}^{k} = && \Omega_k
 \lbrack f_{k}^2(n-k,n_{r})  \frac{n !}{(n -k)!} - f_{k}^2(n ,n_{r})
\frac{ \left( n +k\right) !}{n !}\rbrack \nonumber \\&&
\label{Rabi}
\end{eqnarray}
are effective Rabi frequencies. In Fig. 3, we plot their
dependence, for the particular case $k=1$ and $\eta=0.15$, with
the CM and relative motion quanta number, $\hat{n}$ and
$\hat{n_r}$. This dependence is relevant when considering
realistic situations for the teleportation protocol.

\begin{figure}[t]
\centerline{\resizebox{0.5\textwidth}{!}{
\includegraphics{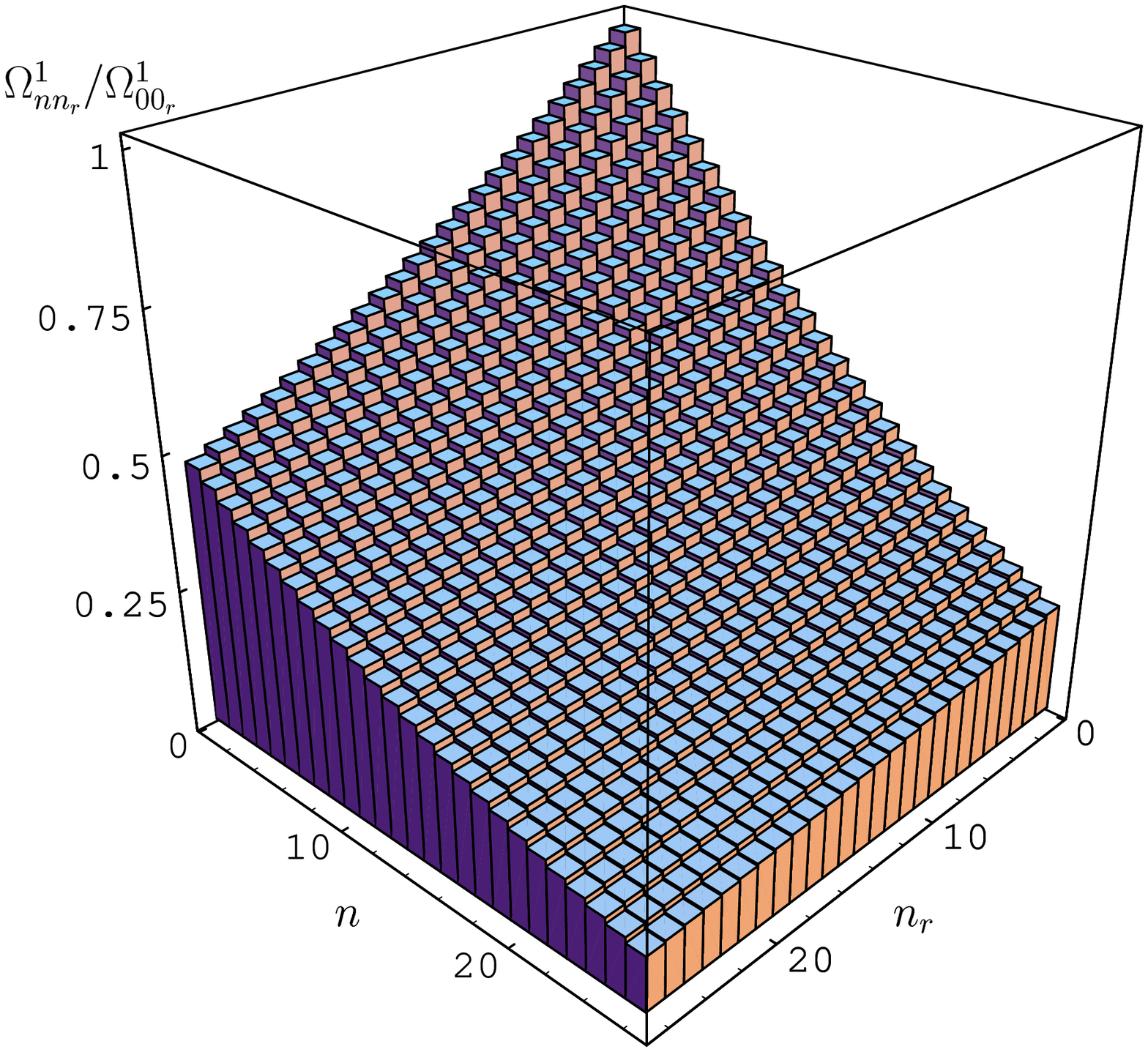}} }
\vspace*{-4cm} \caption{Scaled Rabi frequency $\Omega
_{nn_{r}}^{k}$ plotted as a function of $n$ and $n_{r}$ for $k=1$
and $\eta=0.15$. We stress the fact that we have a different
positive Rabi frequency for each different pair
$(0,0)<(n,n_{r})<(25,25)$. } \label{fig3}
\end{figure}

It is easy to show that in the simpler case $k=1$ and in the
Lamb-Dicke limit the effective Hamiltonian given in
Eq.~($\ref{Heff}$) reduces to
\begin{equation}
H=\hbar \left| \Omega_{1} \right| \left[
S_{+j}S_{+m}e^{2i\phi}-S_{+j}S_{-m}e^{i\phi _{0}}-
\frac{1}{2}\right] +H.c. \label{Hefftelep}
\end{equation}
Notice that, in this limit, this interaction is identical to the
one developped by A. S{\o}rensen and K. M{\o}lmer \cite{sorensen},
who first noticed the relevance of its independence on the
vibrational quantum state of the ions. In the analytical
description of the teleportation scheme we will use this
particular Hamiltonian. Nevertheless, it is clear that when
considering realistic situations in our simulations, as an initial
motional thermal state, we will have to use the full $n$ and $n_r$
dependence given by the more general Hamiltonian presented in Eq.
($\ref{Heff}).$ For the preparation of the Bell state in trap $B$,
we consider that the two ions, $2$ and $3$, were previously cooled
down to the Lamb--Dicke regime and stay in their electronic ground
states. By letting the two Raman laser pairs interact with the
ions during a time $\tau =\pi /(4|\Omega_{1}|),$ we prepare the
Bell electronic state
\begin{equation}
|\Phi_{23}\rangle =\frac{1}{\sqrt{2}}\left( \left| \downarrow
    _{2},\downarrow _{3}\right\rangle
-ie^{2i\phi_B}\left| \uparrow _{2},\uparrow
    _{3}\right\rangle \right),
\end{equation}
where $\phi_B$ is the effective phase of both Raman lasers pairs
in trap $B$. We then carefully transfer ion 2 from trap $B$ to
trap $A,$ where it interacts with ion 1. Possible mechanisms to
make this transfer will be discussed in Section $4$, with the only
necessary, and feasible, condition that no entanglement should
occur between the electronic and motional states. We assume that
after the transfer process the mean quantum number $\bar n$
associated to the vibrational modes will be small, so that the
Lamb--Dicke approximation, $\bar n\eta^2\ll 1$, remains valid in
both traps. If the transfer process itself causes excessive
heating, sympathetic cooling techniques may be used to reduce
$\bar n$. The total electronic state of ions 1, 2 and 3 is then
given by
\begin{equation}
|\Psi_0(1,2,3)\rangle = (\alpha \left| \downarrow
_{1}\right\rangle +\beta \left| \uparrow _{1}\right\rangle )\left|
\Phi_{23}\right\rangle.
\end{equation}

Now we want to consider, in a more realistic way, the generation of
the maximally entangled state. For this purpose, we estimate the
fidelity in the generation of the Bell state $\rho=|\Phi^{(+)}\rangle
\langle \Phi^{(+)} |$ when both ions are initially in their
electronic ground state and in a collective two mode motional thermal
state. By making use of Eq.~(\ref{Evols}) together with
Eq.~(\ref{Heff}), the fidelity $F(\tau)={\rm Tr}\{\rho\rho'(\tau)\}$
of the resulting state $\rho'(\tau)$, after an interaction
time $\tau$, is found to be
\begin{equation}
  \label{fidel}
  {\rm F}(\tau)=\frac{1}{2}+\frac{1}{2}\sum_{nn_r}P_{nn_r}\
\sin{\left(2|\Omega^1_{nn_r}|\tau\right)},
\end{equation}
where $P{nn_r}$ is the initial two mode vibrational thermal
distribution.  We consider an exact $\pi/4-$pulse associated with the
effective Rabi frequency $\Omega _{00_{r}}^{1}$ given in
Eq.~(\ref{Rabi}), so that $\tau=\pi/(4\Omega _{00_{r}}^{1})$.  For the
cases where the vibrational modes are initially in a thermal
distribution with mean number occupation $\bar{n}=0.2,1,5$ and,
considering the same associated temperature,
$\bar{n}_r=0.047,0.43,2.69$, the fidelity given by Eq.~(\ref{fidel})
is $F=0.999,0.988,0.880$, respectively. This high fidelity in the
generation of the quantum channel permit us to continue confidently
with the teleportation protocol.

  It is important to mention that in a recent work \cite{sorensen0}
  S{\o}rensen and M{\o}lmer develop a careful study of the fidelity in
  the generation of a Bell state. Even if they use an alternative
  interaction regime, in the search of a faster operation, we find
  relevant to mention that our fidelity estimations are similar. For
  example, under the same realistic conditions metioned above ($\bar
  n=0.2$ and $\eta=0.2$), their analytical expresions give $F=0.999$,
  while it goes down to $F=0.9$ when $\bar n=5$ and the same
  $\eta=0.2$.  This alternative scheme was already used for generating
  a Bell state in the NIST group \cite{sackett}, providing a first
  fundamental step for the implementation of a teleportation protocol
  in trapped ions.

\subsection{Bell analyzer}

If we follow closely the recipe of the BBCJPW
protocol~\cite{bennett}, our second task would be to perform a
complete measurement, of the von Neumann type, on the subsystem of
particles 1 and 2 in the Bell operator basis, that will
confidently provide two bits of classical information. We propose
here a practical way of obtaining this classical information by
suitably entangling the electronic state of particles 1 and 2 and
then measuring the individual electronic state of each ion. This
substitutes the non-trivial requirement of a direct Bell
measurement by that of monitoring the individual ion fluorescence.
For achieving this goal, we first apply, on ions 1 and 2, a pulse
of duration $\tau=\pi /(4|\Omega_{1} |) $, using the same
excitation scheme as  used before in trap $B$, transforming the
total electronic state $ |\Psi_0(1,2,3)\rangle$ into
\begin{eqnarray}
|\Psi(1,2,3)\rangle = \, \, \, \, \nonumber \\
\frac{1}{2}\Big\{
-ie^{2i\phi_A}\left| \uparrow_1\uparrow_2\right\rangle &&\otimes
\left[ \alpha \left| \downarrow_3\right\rangle +e^{2i(\phi_B-\phi_A)}\beta \left|
\uparrow_3\right\rangle \right] \nonumber \\
+\left| \downarrow_1\downarrow_2\right\rangle &&\otimes
\left[ \alpha \left| \downarrow_3\right\rangle -e^{2i(\phi_B -\phi_A)}\beta \left|\uparrow_3
\right\rangle \right] \nonumber \\
+\left| \uparrow_1\downarrow_2\right\rangle &&\otimes \left[\beta
  \left|
\downarrow_3\right\rangle +e^{i(2\phi_B+\phi_0)}\alpha \left| \uparrow_3\right\rangle
\right] \nonumber \\
+ie^{-i\phi_0}\left| \downarrow_1\uparrow_2\right\rangle &&\otimes
\left[ \beta \left|\downarrow_3\right\rangle -e^{i(2\phi_B+\phi_0)}\alpha \left|
    \uparrow_3
\right\rangle \right] \Big\},
\label{analyzer}
\end{eqnarray}
where $\phi_A$ is the effective phase of both Raman laser pairs in
trap $A$ and $\phi_0$ is the phase due to the equilibrium
separation of the ions 1 and 2 in trap $A$.  If we set, for
simplicity, $\phi_A=\phi_B=\pi-\phi_0/2,$ it is clear that the
determination of the energy state of ions 1 and 2 projects ion 3
on one of the 4 states $\alpha \left| \downarrow_3\right\rangle
+\beta \left|\uparrow_3\right\rangle,$ $\alpha \left|
\downarrow_3\right\rangle -\beta \left|\uparrow_3\right\rangle,$
$\beta \left|\downarrow_3\right\rangle +\alpha \left|
\uparrow_3\right\rangle,$ and $\beta
\left|\downarrow_3\right\rangle -\alpha \left|
\uparrow_3\right\rangle$. These four states corresponds to the
original state or a $\pi$ rotation of it around the $z$ axis, the
$x$ axis or the $y$ axis, respectively. If now Alice measures the
electronic energy of each ion in her trap and send the result (2
bits of classical information) to Bob, he will need or not to make
an additional operation (the corresponding inverse rotation around
the $z$, $x$ or $y$ axis) on ion 3. This completes the
teleportation of the original arbitrary state from particle 1 to
particle 3. We remark that this scheme provides reliable
teleportation \cite{vaidman}, where we have theoretically $100\%$
of success in distinguishing the four Bell states in our analyzer.
This is also the case for atom-cavity systems, as in the proposal
discussed in Ref. \cite{davidovich}, but impossible for photons in
the absence of an effective photon-photon interaction.

It is important to note that, initially, ions 1 and 2 were not correlated and
that the applied pulse entangles these particles, communicating ion 1 with the
quantum channel already established in ions 2 and 3. This is not in
contradiction with the fact that this pulse, formally, transforms unitarily
the Bell basis into a completely disentangled one, as we may see in
Eq.~(\ref{analyzer}), for the proper work of our Bell analyzer.

It is known that the quantum channel and a Bell analyzer in a
teleportation scheme may be implemented using a product of a C-NOT
gate and a Hadamard rotation \cite{preskill}, as it is shown in
Fig.~4.  Using ionic individual addressing, C-NOT gates in the two
ion system have been proposed by Cirac and Zoller~\cite{cirac1},
using $4$ laser pulses on a system of ions cooled to the
vibrational ground state. In the case of thermal ions, S{\o}rensen
and M{\o}lmer~\cite{sorensen} have shown that C-NOT gates may be
implemented with $7$ laser pulses applied on individual ions plus
$2$ laser pulses applied simultaneously on two ions.
The operation equivalent to the product of these two gates  is
realized physically in a simpler
way by means of the interaction in Eq.~(\ref{Hefftelep}). It
realizes {\it simultaneously} what the quantum circuit of Fig.~4
shows as a sequence of two operations.

\begin{figure}[t]
\centerline{\resizebox{0.48\textwidth}{!}{
\includegraphics{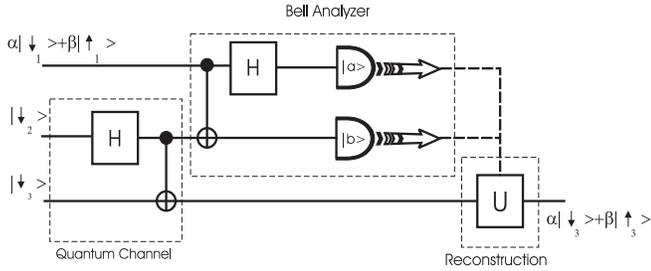}} }
\vspace{0.5cm} \caption{Reliable teleportation quantum logic
circuit associated with a typical teleportation scheme. The
detectors may produce the following two-cbit information,
$|a\rangle=\lbrace|\downarrow_1\rangle,|\uparrow_1\rangle \rbrace$
and $|b\rangle=\lbrace|\downarrow_2\rangle,|\uparrow_2\rangle
\rbrace$, that will condition Bob's $U$ rotation to reconstruct
the teleported state. In our scheme each operation, ${\rm
H}\otimes {\rm C\!\!-\!\!NOT}$ or ${\rm C\!\!-\!\!NOT}\otimes{\rm
H},$ is substituted by a single laser pulse.  } \label{fig4}
\end{figure}

\subsection{Teleportation fidelity}

We study some possible sources of error when considering the
teleportation process, say, imprecision in the lasers pulse area
and the effects of finite temperature in the initial vibrational
state in both traps. We start by considering an ideally well
generated Bell state in ions 2 and 3 in trap B, due to the good
results of simulations under realistic conditions discussed above.
After ion 2 is transported adiabatically from trap B to trap A,
leaving ion 3 alone in trap B, the two traps thermalize in such a
way that we may consider a collective two mode thermal state in
trap A and a one mode motional thermal state in trap B. When associating a
common temperature to both collective modes in trap A, the
relation between the CM phonon mean value, $\bar{n}$, and the
relative phonon mean value, $\bar{n_r}$, is
\begin{equation}
\bar{n}_r=\frac{1}{(\frac{1}{\bar{n}}+1)^{\frac{\nu_r}{\nu}}-1} ,
\end{equation}
where $\nu_r/\nu=\sqrt{3}$ in our case. For simplicity, we will
assume that trap B thermalizes to the same temperature than trap
A. We will apply now the required $\pi/4-pulse$ on ions 1 and 2,
in trap A, with the effective Hamiltonian of Eq.~(\ref{Heff}). We
choose the time of action of the pulse to be
$(1+\epsilon)\pi/(4|\Omega^{1}_{0,0}|)$, where $\epsilon$ is an
imprecision factor in the pulse area and $\Omega^{1}_{0,0}$ is the
effective Rabi frequency for the collective vibrational ground
state. Clearly, for the considered initial density matrix and for
$\epsilon=0$, only the terms associated with the vibrational
fundamental mode will evolve, as desired, from the Bell state
basis to the completely decoupled basis. The other Fock states in
the thermal distribution of trap A will produce  different final
states, given rise to an unavoidable error source. We measure now
the global state of ions 1 and 2 in the disentangled basis. Then,
ion 3 is left, up to an eventual local unitary operation depending
on the result of the measurement, in the original state of ion 1.
This last operation on ion 3 will introduce the same kind of
errors as before, due to the one mode thermal distribution in trap
B and the imprecision in the lasers pulse area.

An estimation of teleportation fidelity may be done, as it is
shown in Fig. 5, using the relation
$F=Tr\lbrace\rho_3\rho'_3\rbrace$, where $\rho_3$ stands for the
ideal teleported state in ion 3 and $\rho'_3$ for the realistic
generated state. In Fig. 5(a), no imprecisions in the lasers pulse
area are considered, showing only the effect of the increasing
$\bar{n}$ and $\eta$. Fidelity $F$ goes down from unity, under
ideal conditions of a pure vibrational ground state and Lamb-Dicke
limit, to around $0.985$, for larger collective phonon thermal
states and higher Lamb-Dicke parameters. In Fig. 5(b), we add a
rather large imprecision in the lasers pulse area $(5\%)$ and,
nevertheless, obtain a Fidelity $F$ which is never less than
$0.975$. We recall that in very recent experiments in the coherent
manipulation of two ions \cite{king}, typical values of
$\bar{n}=0.11$ and $\bar{n_r}=0.01$ are reported. We may say,
then, that our teleportation protocol is robust under realistic
conditions studied here.
\begin{figure}[htb]
\centerline{\resizebox{0.67\textwidth}{!}{
\includegraphics{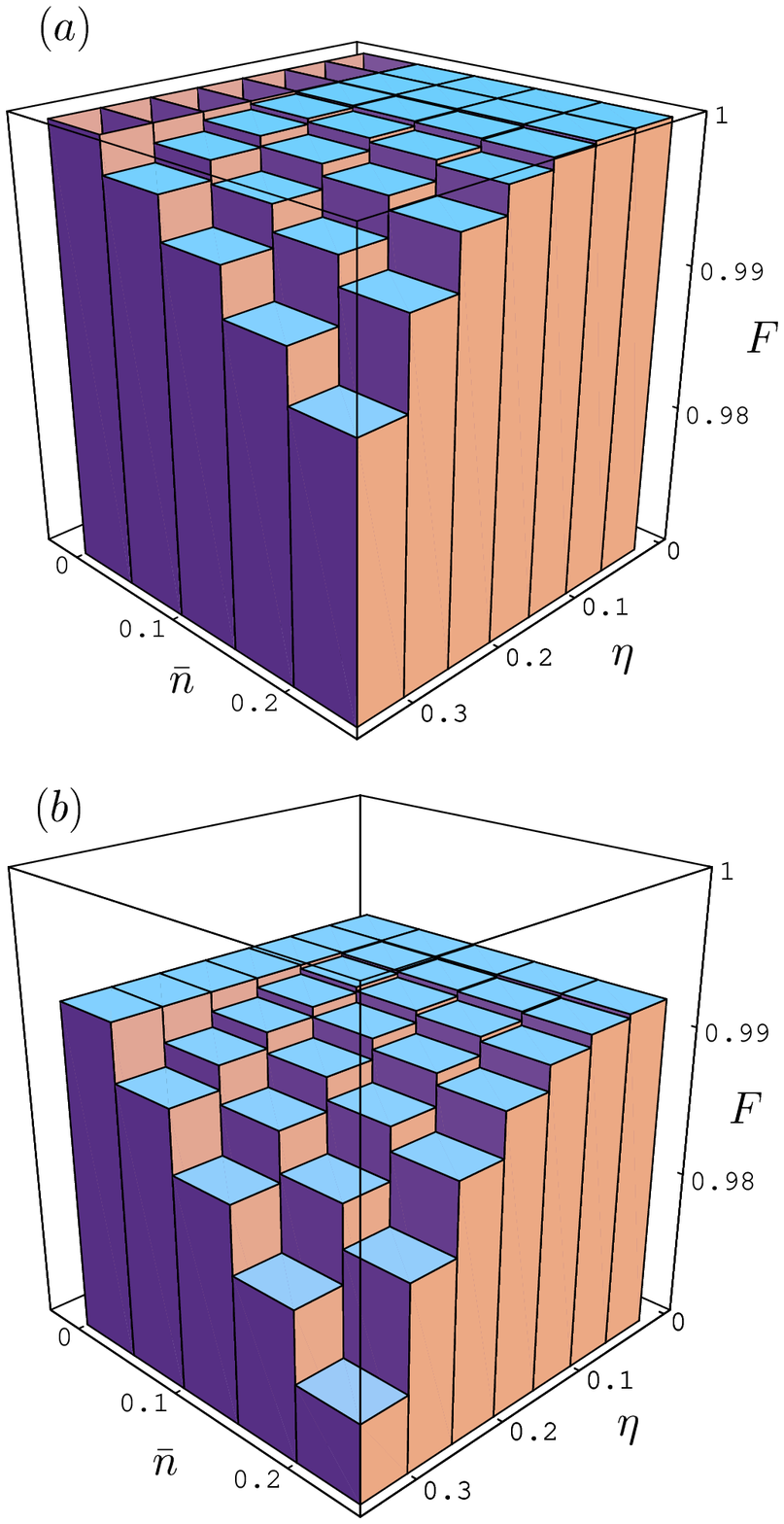}} }
\vspace*{-4cm} \caption{Fidelity of the teleported state,
$F=Tr\lbrace\rho_3\rho'_3\rbrace$, as a function of the motional
phonon mean number $\bar{n}$ and the CM Lamb-Dicke parameter
$\eta$, considering $0\%$ (a) and $5\%$ (b) of imprecision in the
pulse area, respectively.} \label{fig5}
\end{figure}

\subsection{BBCJPW technicalities}

We would want to discuss now some considerations about the standard
teleportation protocol \cite{bennett}. The principal scope of our scheme is to
{\it test} reliable teleportation in massive particles. For achieving this
goal, some minor technicalities, as if it is easier or not to send the
particle instead of teleporting its quantum state, are of no relevance. In
fact, if we think in {\it tests} of teleportation of photonic polarization
states, it would be even easier and quicker to send the photon. In this way,
we may argue that the experimental realization of teleportation of internal
states in ions may be considered as a fundamental step in the coherent
manipulation of quantum information and, for sure, an important task in the
implementation of a quantum channel between quantum networks. On the other
hand, our proposal considers and respects the importance of the arbitrariness
of the state in particle 1 as claimed in the standard teleportation protocol,
as long as the teleporting scheme must be universal in the sense of state
independent. However, it is of no relevance if a third person (usually called
Victor) prepares in private an arbitrary state in particle 1 and gives it to
Alice, or if Victor prepares the arbitrary state close to Alice, as she may
"close her eyes" politely. What should remain clear, as we have stressed
before, is that the implementation of our proposal would lead to {\it reliable
  teleportation}, fact that seems impossible for photonic polarization states
in the absence of photon-photon interaction.

\section{entanglement teleportation and entanglement swapping}

We discuss, briefly and without entering into technical details, natural
extensions of our teleportation scheme, i.e., teleportation of an entangled
state \cite{bennett} and the so-called entanglement swapping \cite{pan}. For
the case of teleportation of an entangled state we need initially two pairs of
ions, each pair in a different trap. The first pair in trap $A$, ions 1 and 2,
is in an arbitrary unknown entangled state, and the second pair, ions 3 and 4,
must be in a Bell state. First of all, we transport adiabatically ion 1 to
another trap, preserving the entanglement in the internal degrees of freedom.
Then, we apply to ions 2, 3 and 4 the same operations used previously for
single state teleportation to ions 1, 2 and 3, resulting in the transfer of
the initial arbitrary entanglement of ions 1 and 2 to the ions 1 and 4. For
achieving entanglement swapping, we consider two traps containing, each one, a
pair of ions in a Bell state. For example, we may consider the total initial
state $|\Phi^{+}\rangle_{12} |\Phi^{+}\rangle_{34}$, that may be rewritten as
\begin{eqnarray}
\frac{1}{2} \lbrace |\Phi^{+}\rangle_{14} |\Phi^{+}\rangle_{23}+
|\Phi^{-}\rangle_{14} |\Phi^{-}\rangle_{23} \nonumber \\
+|\Psi^{+}\rangle_{14} |\Psi^{+}\rangle_{23}+
|\Psi^{-}\rangle_{14} |\Psi^{-}\rangle_{23} \rbrace .
\end{eqnarray}
From this equation it is clear that if we project ions 2 and 3 onto any of the
four Bell states, we create a similar Bell state between ions 1 and 4, in
spite of the fact that these particles never interacted before. To this end,
we will take advantage of the ionic internal state Bell analyzer. For
distinguishing the four Bell states in ions 2 and 3 we put this pair into an
individual trap, preserving their initial internal correlations with isolated
and distant particles 1 and 4, respectively. Then, we apply on ions 2 and 3 a
pulse $\pi/4$ with the interaction described by the Hamiltonian in Eq.
(\ref{Hefftelep}), producing the state
\begin{eqnarray}
\frac{1}{2} \lbrace |\Phi^{+}\rangle_{14} |\uparrow\uparrow\rangle_{23}+
|\Phi^{-}\rangle_{14} |\downarrow\downarrow\rangle_{23} \nonumber \\
+|\Psi^{+}\rangle_{14} |\uparrow\downarrow\rangle_{23}+
|\Psi^{-}\rangle_{14} |\downarrow\uparrow\rangle_{23} \rbrace .
\end{eqnarray}
Hereafter, we just have to measure the individual internal state of ions 2 and
3 to project particles 1 and 4 onto the corresponding Bell state. Again, we
have a theoretical $100\%$ of success in the process. This was also the case
in the previously described teleportation scheme, but different from photonic
entanglement swapping proposals~\cite{pan}.

In our view, technical tasks to implement the present teleportation scheme,
that will be discussed in the next Section, are similar to those for
entanglement teleportation and entanglement swapping. In this sense, we
believe that these and other experimental challenges in trapped ions may be
envisaged.

\section{Experimental issues}

We now briefly discuss experimental issues, proposing one scenario in which
our teleportation scheme could be implemented.
While many of the required tasks have been individually demonstrated, others
are being actively pursued by experimental groups. For this discussion we
borrow heavily on the implementations ideas of Wineland and
co-workers\cite{review}, as they have been able to successfully address many
of the questions raised in this paper. Below is a list and discussion of the
required experimental steps. (i) The deterministic generation of an arbitrary
one ion internal state $ \alpha \left| \downarrow \right\rangle +\beta \left|
  \uparrow \right\rangle $ has already been achieved (see for example
Ref.~\cite{meekhof}). (ii) Ions may be cooled down to such low
temperatures that, in a good approximation, we may consider
initial density operators of the form $\left| \downarrow
\downarrow \right\rangle \langle \downarrow \downarrow\!\!|\otimes
\hat{\rho} _{vib}, $ with $\hat{\rho} _{vib}$ associated with a
distribution were the vibrational ground state is heavily
populated\cite{king}. (iii) The deterministic generation of a Bell
state and the disentangling of the Bell basis, outlined above, may
be implemented with state of the art technology through the
procedure described in detail above. The results of the numerical
simulations show that we may be optimistic about the experimental
realization of ionic internal Bell states.  (iv) Transporting ions
from one trap to another requires the construction of novel traps.
The envisioned traps would have lithographically deposited
electrodes~\cite{review} for controlling the displacement of the
ions without affecting the internal states. Such traps are being
pursued by the group at NIST \cite{wineland}. These traps should
allow for a smooth, almost adiabatic, separation of ions from a
single potential well, and the reverse operation, without much
heating to the ionic motion.  Also, dynamic electric and magnetic
transport would not be out of question, so long as the heating is
not large and the internal states remain untouched. For that, the
switching of electric fields are to be made adiabatic with respect
to the internal states, though not necessarily adiabatic with
respect to the ions vibrational frequency. Since the theoretical
method above for the generation of entanglement is nearly
independent of the vibrational quantum number, in the Lamb--Dicke
regime, one only has to worry about large heating mechanisms. If
such heating exists, it can still be overcome, at added
complexity, through the use of sympathetic cooling to cool again
the joined ions to low vibrational states, without affecting the
relevant ions internal state. For example, sympathetic cooling of
ions in a linear Paul trap is in the scope of W. Lange and
co-workers \cite{lange} to realize quantum computing experiments.
In the same row, magnesium ions, suitable for doing quantum logic
operations, will be mixed with indium ions, suitable for
sympathetic (indirect) cooling of the ions communicated by a
collective mode. Recently, J. Hangst and co-workers in Aarhus have
performed beautiful experiments demonstrating sympathetic cooling
crystallization of trapped ions \cite{bowe}. They were able to
keep up to 14 ordered cooled ions by the direct cooling on a
single $^{24}$Mg$^+$ ion. (v) The production of the pairs of Raman
laser beams and the pulse shaping (power and duration) can be
accomplished with the proper use of acousto-optical modulators and
RF switches, and the relative phases can be controlled by the path
lengths. Laser power levels and timing, while dependent on a
chosen experimental system, are estimated to be compatible with
what is currently used for Raman excitation of trapped ions (as
for example, in Ref.\cite{king}). (vi) Finally, the readout of the
internal states of the individual ions 1 and 2 is most easily done
by separating them, using the same transport mechanism as above,
and using the "electron-shelving" technique~\cite{dehmelt} for
achieving nearly 100\% detection efficiency in each resolved ion.
An alternative method, not requiring a separation of the ions,
would follow Leibfried's~\cite{leibfried} proposal to read out
individual electronic states of two or more trapped ions, selected
by their rf-micromotion. We also note the work done in R. Blatt's
group \cite{nagerl} in Innsbruck, where they were able to directly
address individual ions, in an optical transition, by shifting
either the ions string or the laser beam with acousto-optical
modulators. Unfortunately, their scheme is not suitable for the
other tasks required for our proposal as they are in a "weak trap"
regime, not being able to perform sideband cooling towards the
ground state. More recently, the same group has used a stronger
trap and achieved sideband cooling to the ground state on a {\it
single} Ca+ ion \cite{roos}. (vii) Of course, states $|\!\uparrow
\rangle$ and $|\!\downarrow \rangle$ are either different
hyperfine states in the proper magnetic field, such that their
energy's difference is field independent, or different metastable
electronic states \cite{roos,king}, making this a robust system
with a long coherence time. The specific experimental system will
dictate the requirements on laser and trap stabilities, among
other experimental parameters.

While the experimental scenario envisioned above is not readily available, we
believe it soon will be, as inferred from the latest impressive developments
in trapped ion technology.

All arguments presented above convince us that teleportation with massive
particles, not speaking about its obvious fundamental interest, are achievable
and interesting experimental tasks that will result in several applications in
the burgeoning field of quantum information.

\section{conclusions}

In summary, we have shown that a reliable teleportation scheme is
possible for the internal states of trapped ions and, to this end,
we have proposed a suitable Bell analyzer based on an operation
that transforms unitarily the Bell basis into a disentangled one.
We have presented numerical simulations to show the influence of
the technical imprecisions and finite temperature on the desired
result. Entanglement teleportation and entanglement swapping are
briefly discussed, as they are in fact natural extensions of state
teleportation. Finally, a step by step discussion of a probable
experimental realization of ionic internal Bell states and
teleportation is presented.

\begin{acknowledgement}
This work was partially supported by the Conselho Nacional de
Desenvolvimento Cient{\'\i}fico e Tecnol\'ogico (CNPq),
Funda\c{c}\~ao de Amparo \`a Pesquisa do Estado do Rio de Janeiro
(FAPERJ) and the Programa de Apoio a N\'ucleos de Excel\^encia
(PRONEX). We acknowledge useful information from D. Wineland
concerning experimental issues, particularly on the implementation
of lithographic traps. E. S. would like to thank hospitality and
useful discussions, especially about sympathetic cooling and the
implementation of our teleportation scheme, to W. Lange and S.
Koehler at Max-Planck Institut (Garching, Germany).
\end{acknowledgement}

{}

\end{document}